\documentclass[aps,prl,twocolumn,showpacs,showkeys,nofootinbib,superscriptaddress]{revtex4-1}

\usepackage[utf8]{inputenc}
\usepackage{graphicx}
\usepackage{amsfonts}
\usepackage{amssymb}
\usepackage{amsmath}
\usepackage[mathscr]{eucal}
\usepackage{setspace}
\usepackage{booktabs}
\usepackage[shellescape,latex]{gmp}
\usempxpackage{amssymb}

\def\sqr#1#2{{\vcenter{\hrule height.#2pt
   \hbox{\vrule width.#2pt height#1pt \kern#1pt
      \vrule width.#2pt}
   \hrule height.#2pt}}}

\def\bsqr#1#2{{\vrule width #1pt height#2pt}}
\def\bsquare{{\mathchoice\bsqr66\bsqr66\bsqr33\bsqr33}}
\def\badbreak{\penalty1000}

\def\fir{{\scriptscriptstyle{\text{\rm IR}}}}                 
\def\lm0{{\lambda_0}}                                             
\def\nrN{N}                                                       
\def\cf{\mathfrak{n}}                                         
\def\cfu{\cf_\star}                                              
\def\efN{\mathscr{N}}                                        
\def\efNm{\efN_\star}                                        
\def\w{c}                                                              
\def\v{b}                                                               

\def\Veff{V_{\text{eff}}}

\newcommand*{\LessApprox}{\smallrel\lessapprox}

\makeatletter
\newcommand*{\smallrel}[2][.8]{%
  \mathrel{\mathpalette{\smallrel@{#1}}{#2}}%
}
\newcommand*{\smallrel@}[3]{%
  \sbox0{$#2\vcenter{}$}%
  \dimen@=\ht0 %
  \raise\dimen@\hbox{%
    \scalebox{#1}{%
      \raise-\dimen@\hbox{$#2#3\m@th$}%
    }%
  }%
}
\makeatother

\usepackage{amsmath} 
\usepackage{dsfont}  
\usepackage{bm}      

\def\beq{\begin{equation}}
\def\eeq{\end{equation}}
\def\beqs#1\eeqs{\beq\begin{split} #1 \end{split}\eeq}

\long\def\comment#1{}

\def\be{\begin{equation}}
\def\ee{\end{equation}}

\def\bc{\begin{center}}
\def\ec{\end{center}}

\def\corr#1{\textcolor{red}{#1}}

\def\sdiag{\sigma_{\rm diag}}

\usepackage{microtype}
\usepackage[dvipsnames]{xcolor}
\usepackage[colorlinks=true,backref=false, linktocpage=true,
citecolor=PineGreen,urlcolor=PineGreen,linkcolor=PineGreen,pdfpagemode=UseOutlines]{hyperref}

\hypersetup{%
  bookmarksnumbered=true,
  pdftitle = {},
  pdfsubject = {},
  pdfauthor = {},
  pdfkeywords = {}
}

\begin{document}

\title{Super-Universality in Anderson Localization}

\author{Ivan Horv\'ath}
\email{ihorv2@g.uky.edu}
\affiliation{University of Kentucky, Lexington, KY 40506, USA}
\affiliation{Nuclear Physics Institute CAS, 25068 \v{R}e\v{z} (Prague), Czech Republic}

\author{Peter\ Marko\v s}
\email{peter.markos@fmph.uniba.sk}
\affiliation{Dept. of Experimental Physics, Faculty of Mathematics, 
Physics and Informatics, Comenius University in Bratislava, Mlynsk\'a Dolina 2, 
842 28 Bratislava, Slovakia}

\date{Aug 19, 2022}

\begin{abstract}

We calculate the effective spatial dimension $d_\fir$ of electron modes at 
critical points of 3D Anderson models in various universality classes (O,U,S,AIII). 
The results are equal within errors, and suggest the super-universal value 
$d_\fir \!=\! 2.665(3) \!\approx\! 8/3$. The existence of such a unique marker
may help identify natural processes driven by Anderson localization, 
and provide new insight into the spatial geometry of Anderson transitions. 
The recently introduced $d_\fir$ is a measure-based dimension of 
Minkowski/Hausdorff type, designed to characterize probability-induced 
effective subsets.

\medskip

\keywords{Anderson localization, effective dimension, universality, symmetry, 
          effective number, measure}

\end{abstract}

\maketitle


\noindent
{\bf 1.~Introduction. $\,$}
Localization of quantum particles~\cite{Anderson:1958a} is an important
effect influencing the transport properties of mesoscopic systems. 
Following a 1-parameter scaling theory~\cite{Abrahams:1979a}, the transition 
from extended (metallic) to localized (insulating) state takes place at 
a critical point. In a prototypical Anderson model realization, this occurs 
on a critical line in $(E,W)$ plane, where $E$ is the Fermi energy and $W$ 
the strength of a random potential. After years of theoretical and numerical 
investigation (see e.g.~\cite{abrahams201050}) it is generally accepted that 
these Anderson transitions are universal. There are ten universality 
classes~\cite{Evers_2008} with distinct values of critical exponents $\nu$ 
and $s$ describing the approach to criticality from the localized and 
extended sides respectively. Given that these indices are 
coupled~\cite{Wegner:1976a}, 
many numerical works evaluated $\nu$ and the value $W_c$ of disorder at 
the canonical critical point $(0,W_c)$, especially for the 3D orthogonal
class (see e.g.~\cite{MacKinnon:1981a,Slevin:1999a,Slevin_2018}). 

Among key attributes of a transition to localized state is that it 
drastically reduces the volume effectively accessible by a particle. 
Here we describe this effect in a meaningful quantitative manner. 
Note that it is not the endpoint of the Anderson transition that is 
interesting in this regard. Indeed, the known feature of exponentially 
bounded wave function (exponential localization) provides the relevant 
information in that case. Rather, it is the intermediate step toward 
localization, the critical state, that is of primary interest here. 
Indeed, we will describe Anderson criticality in terms of particle's 
propensity to occupy the volume of space nominally available to it.

At the heart of such description is the notion of effective volume.
Since ordinary volume is an important physical attribute of a system, 
so is its effective counterpart if it can be put on a similar conceptual 
footing. In other words, if it can be interpreted as a geometric 
characteristic expressing the amount of occupied space (its measure), 
and is not too arbitrary so as to be uninformative. Such issues have 
recently been successfully 
resolved~\cite{Horvath:2018aap,Horvath:2018xgx,Alexandru:2021pap}. 
Here we will use these results, 
reviewed in Sec.2, to work with properly defined effective volumes of 
Anderson eigenmodes. 

A robust characteristic of a critical point needs to involve 
thermodynamic ($L \to \infty$) limit in order to capture its 
non-analyticity. Thus, 
consider Anderson lattice system in $D$ dimensions so that its volume 
$V(L)\propto L^D$. The effective volume $\Veff(L,E)\le V(L)$ 
occupied by electron at energy $E$ may scale differently,
namely $\Veff(L,E) \propto L^{d_\fir(E)}$, for $L \to \infty$, 
where $d_\fir(E) \!\le\! D$. The {\em effective dimension} $d_\fir$, first 
used in the context of QCD Dirac eigenmodes~\cite{Alexandru:2021pap}, 
properly quantifies the property we seek. Note that, if $d_\fir \!<\! D$, 
the particle occupies space of measure zero relative to nominal space in 
$L \to \infty$ limit ($\Veff/V \rightarrow 0$). The value 
$D \!-\! d_\fir$ gives the rate at which modes lose the ability 
to fill the growing space.

Dimension $d_\fir$ is a strictly infrared (IR) quantity since 
it encodes the asymptotic response of effective volume to the change of 
maximal distance available~\cite{Alexandru:2021pap}. Hence, it 
is a natural characteristic of criticality that transforms spatial
features. In that vein, $d_\fir$ is expected to be universal 
in Anderson transitions, even though it doesn't enter the standard 
scaling theory~\cite{Abrahams:1979a}. However, in this Letter we 
present evidence for something unexpected, namely that 
$d_\fir$ is in fact {\em super-universal}: it expresses commonality 
existing across the symmetry classes. More concretely, we find that 
$d_\fir \!\approx\! 8/3$ for classes O, U, S and AIII, with errors 
(couple parts per mill) comparable to their mutual differences. 
The stark difference between the usual universality (via $\nu$) and 
the proposed super-universality (via $d_\fir$) can be seen from 
the comparison shown in Table~I.


\noindent
{\bf 2.~Effective Volume and IR Dimension. $\,$}
Volume $V \!=\! a^D \nrN$ of a hypercubic system with lattice constant 
$a$ can be thought of as determined by counting the lattice 
sites $(\nrN)$. 
Its measure-like nature then stems from additivity of ordinary 
counting: the total is the sum of counts for parts.
Similarly, the effective volume $\Veff[\psi] \!=\! a^D \efN[\psi]$ occupied 
by wave function $\psi$ is determined by effective counting $(\efN)$ which 
takes into account that $\psi$ endows lattice sites with probabilities, and 
hence varied relevance. For $\Veff$ to be measure-like, the underlying 
effective counting also has to be additive. 

The effective number theory of Ref.~\cite{Horvath:2018aap} determines 
all consistent additive schemes to count collections of objects 
$(o_1,o_2,\ldots, o_\nrN)$ with probabilities 
$P\!=\!(p_1,p_2,\ldots, p_\nrN)$. Each scheme is represented by function 
$\efN \!=\! \efN[P]$ on discrete probability distributions. A key result 
is the existence of a scheme $\efNm$ satisfying 
$\efNm[P] \le \efN[P]$ for all $P$ and $\efN$. This minimal effective 
amount, specified by
\begin{equation}
      \efNm[P]  \,=\, \sum_{i=1}^\nrN \cfu(\nrN p_i)   \quad,\quad
      \cfu(\w)  \; = \;   \min\, \{ \w, 1 \}    \;
      \label{eq:020}         
\end{equation}
is inherent to each collection\footnote{It is inherent because 
it cannot be reduced by redefinition of a counting scheme. When applied 
to uncertainty in quantum mechanics, this leads to the notion of intrinsic 
uncertainty~\cite{Horvath:2018xgx}.}, and has absolute meaning.

Effective dimension $d_\fir$ is based on 
$\efNm$~\cite{Horvath:2018aap,Alexandru:2021pap,Horvath:2022ewv}.
The explicit definition for Anderson systems starts with eigenfunction
$\psi \!=\! \psi(x_i,E,W,L,\zeta)$ at a particular realization $\zeta$ 
of microscopic disorder with strength $W$. The associated probabilities 
are $p_i \!=\! \psi^+ \psi(x_i)$, and $d_\fir(E,W)$ arises via
\begin{equation}
     \langle \, \efNm \,\rangle_{E,W,L} \,\propto\, L^{d_\fir(E,W)} 
     \quad\; \text{for} \quad\; L \to \infty  \quad 
     \label{eq:040}	
\end{equation} 
where $\langle\dots\rangle$ is the disorder average involving states 
from the spectral vicinity of $E$. Here we will be interested in
$d_\fir(0,W_c)$ for models from O,U,S and AIII symmetry classes. 
Their critical values $W_c$, listed in Table~I, are known to high 
accuracy.

\begin{table}[b!]
	\begin{tabular}{|c|l|l|l|l|l|}
		\hline
		model & $\,$Ref$\,$ & $\;W_c\;$ & $\nu$ & $d_\fir$(here) \\
		\hline
		O   & $\,$\cite{Slevin_2018}  & 16.543(2)   & 1.572(5) & 2.664(2) \\
		U   & $\,$\cite{Slevin:1999a} & 18.375(17)  & 1.43(6) &  2.665(3)  \\
		S   & $\,$\cite{Asada_2005}   & 19.099(9)   & 1.360(6) & 2.662(4)  \\
		A   & $\,$\cite{Wang_2021}$\,$    & 11.223(20)~ & 1.071(4)~ & 2.668(4) \\
		\hline
	\end{tabular}
	\caption{Critical parameters of 3D orthogonal (O), unitary (U), symplectic (S) 
	and AIII (A) symmetry classes. Their meaning is discussed in the text.}
	\label{table:1}
\end{table}

\noindent
{\bf 3. Anderson Models.}
All models we study are defined on $L^3$ cubic lattice with each site  
$r\!=\!(x,y,z)$ supporting two quantum states. Disorder is introduced 
via random energies $\epsilon_r$ chosen from a box distribution 
in the range $[-W/2,+W/2]$. 
Hopping terms in the Hamiltonian only connect the nearest neighbors. 
They are described by $2\times 2$ matrices $t_{r,e_j}$, one for each 
$r$ and direction specified by the unit lattice vector $e_j$ ($j=x,y,z$).
The Hamiltonian~is
\be
{\cal H} \,=\, \sum_r\epsilon_r \, c^\dag_r \,\sdiag \,c_r 
\,+\, \sum_{r,j}  c^\dag_r \, t_{r,e_j} c_{r-e_j} + h.c.
\ee
where operators $c_r$ have two components and $\sdiag$ is diagonal. 
In definitions of specific models below, $\sigma_0$ denotes 
the identity matrix and $\sigma_j$ the Pauli matrices. Periodic boundary
conditions are imposed in all cases.

\smallskip
\noindent
{\bf Orthogonal} (O)$\,$: $\sdiag = t_{r,e_j} = \sigma_0$.

\smallskip
\noindent
{\bf Unitary} (U)$\,$: $\sdiag \!=\! t_{r,e_x} \!=\! t_{r,e_y} \!= \sigma_0$ 
and $\theta \!=\! 1/4$ \cite{Slevin:1997a} in 
\begin{equation}
t_{r,e_z} \!= \sigma_0\exp{(-i 2\pi\theta x)}  \;
\end{equation}
{\bf Symplectic Ando} (S)$\,$: $\sdiag = \sigma_0$ and 
$\theta \!=\! \pi/6$ \cite{Asada_2005} in
\be
t_{r,e_j} = \exp(i\theta\sigma_j)
\ee
{\bf AIII} (A)$\,$: $\sdiag \!=\! \sigma_z$. We use $t_\parallel \!=\! 0.4$
in $t_{r,e_z} = t_\parallel\sigma_0$, and 
$t_1 \!=\! t_2 \!=\! 0.5$, $t_\perp \!=\! 0.6$~\cite{Wang_2021} in
\be
  t_{r,e_x} \!= t_1\sigma_0+it_\perp\sigma_x \quad , \quad
  t_{r,e_y} \!= t_2\sigma_0+it_\perp\sigma_y 
\ee  
We note that in O, U and S models, all energy eigenvalues are doubly 
degenerate, while in the chiral A model they come in $(-E,E)$ pairs for
each sample of disorder.

\noindent
{\bf 4. Technical Details.}
We use JADAMILU library~\cite{jadamilu_2007} to numerically 
diagonalize 1-particle Anderson Hamiltonians. For each sample of 
disorder, we compute $10$ distinct eigenvalues closest to 
$E\!=\!0$ and all associated eigenstates. This results in probing 
a very small vicinity of the band center for all studied 
systems, e.g. $|E| \LessApprox 5\cdot 10^{-3}$ for O at $L=24$ with 
$L^{-3}$ size dependence. 
Hence, all computed states are included in the estimate (simple average) 
of $\efNm$ associated with a given sample. Disorder average is then 
performed by accumulating $5\!-\!20 \times 10^3$ independent samples. 
The sizes of studied systems range from $L=8$ to $L=128$ (O), 
$112$ (U), and $72$ (S) and (A). 

As an intermediate step toward extracting the dimension $d_\fir$, 
we define its finite-$L$ counterpart from ratios of effective
volumes on systems with sizes $L$ and $L/s$
\begin{equation}
    d_\fir(L,s) \,\equiv\, \frac{1}{\ln{s}} \, 
             \ln \frac{\langle \, \efNm \,\rangle_{L}}
                      {\langle \, \efNm \,\rangle_{L/s}}
    \quad , \quad s > 1
    \label{eq:060}	    
\end{equation}
Given the defining asymptotic behavior \eqref{eq:040}, we can then
use that $d_\fir = \lim_{L \to \infty} d_\fir(L,s)$, independently 
of $s$. The latter can be adjusted to suit the available range of 
sizes and statistics. We will use $s\!=\!2$ which is also convenient 
due to a large number of pairs $(L,L/2)$ accessible by the lattice 
geometry. Note that, since the data at different $L$ are independent, 
the error $\Delta$ of $d_\fir(L,s)$ can be estimated 
via simple error propagation. In particular, 
$\Delta(L,s) = \sqrt{\epsilon^2(L) + \epsilon^2(L/s)}/\ln{s}$, where
$\epsilon$ are relative errors of $\efNm$. 

\begin{figure}[t]
   \bc
   \includegraphics[width=0.48\textwidth]{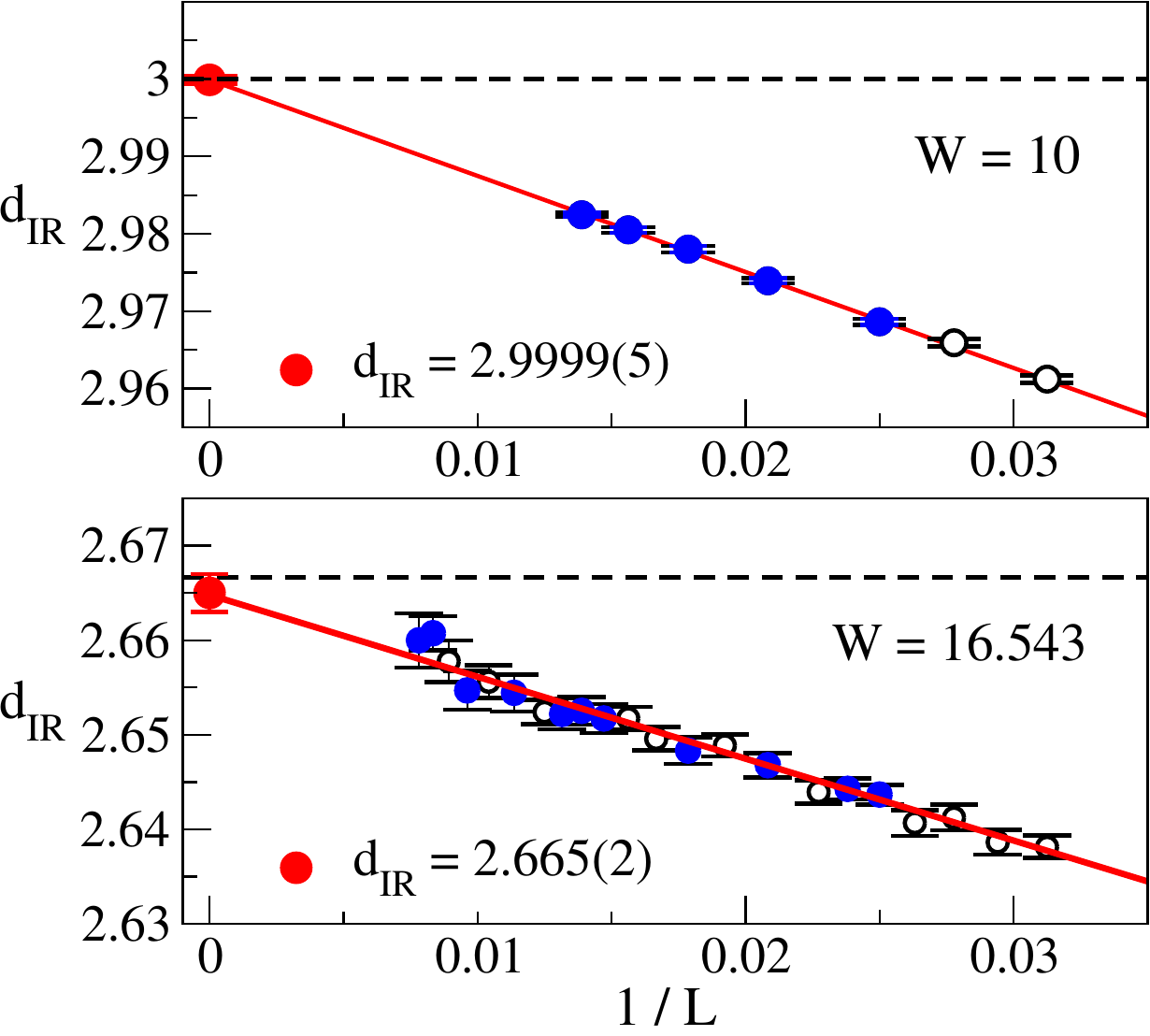}
   \ec
   \vskip -0.15in
   \caption {Sample computation of $d_\fir$ for O system in 
   the metallic regime (top) and at the critical point (bottom).
   Values at finite $L$ are obtained from Eq.~\eqref{eq:060} 
   with $s\!=\!2$. The horizontal dashed line in the bottom 
   plot shows $d_\fir \!=\!8/3$.}
   \label{fig:test}
   \vskip -0.17in
\end{figure}

\smallskip
\noindent
{\bf 5. Sample Computation.}
In order to describe and test our numerical procedure of extracting $d_\fir$, 
we first perform an illustrative calculation in the context of O class. 
In particular, we will evaluate $d_\fir(E\!=\!0,W\!=\!10)$, which is deeply 
in the extended phase (see Table~\ref{table:1}).

To that end, we have generated ensembles for lattice sizes between $L\!=\!16$ 
and $L\!=\!72$. From the set of computed lattices, seven distinct pairs 
$(L/2,L)$ can be formed. We have calculated $d_\fir(L)$ for each of them using 
the relation~\eqref{eq:060} with the result shown in Fig.~\ref{fig:test} (top) 
as a function of $1/L$. A striking feature of the obtained behavior is a clean 
linear approach to the infinite volume limit. However, a direct linear fit 
cannot be used to obtain the extrapolated dimension and its error. Indeed, some 
pairs of points in this graph are correlated since their data input involves 
a common lattice. One way to proceed is to select a suitable subset of mutually 
independent pairs to obtain a valid estimate.

In order to exactly mimic the procedure that will be used to analyze critical 
points, we proceed as follows. We only allow systems of sizes at least
$L_\text{min}\!=\!20$ to participate in the analysis and, given this cut, 
determine the maximal number $K$ of independent pairs $(L/2,L)$ that can be 
formed from the available data. If there is only one maximal combination, 
the associated linear fit determines our final estimate and its error. 
If there are multiple combinations, we quote the average $d_\fir$ over
such determinations and the average error. The ensuing variability 
of the participating estimates characterizes the robustness of the method.

Applying the above to our $W\!=\!10$ data (Fig.~\ref{fig:test}, top) 
yields $K\!=\!5$ and a unique combination of pairs marked in blue. 
Note that the smallest pair is formed by lattices $(20,40)$ so that
we are dealing with the range $1/L \!\le\! 0.025$. The associated fit 
(shown) returns the expected value $d_\fir \!=\! 3$ with the accuracy 
of couple parts in ten thousand. Note that, in this case, the resulting 
fit works extremely well even outside the fitting range implied by 
the size cut. 

Following the same strategy at the critical point, we plot
in Fig.~\ref{fig:test} (bottom) the dimensions $d_\fir(L)$ for
available pairs of sizes. We collected data for 29 lattices satisfying 
the size cut, producing $K \!=\!11$ with 128 distinct combinations. 
One of them is visualized by blue points and the corresponding linear 
fit. The result is close to $d_\fir \!=\! 8/3$ (dashed line) 
and error about one part per mill.

\smallskip
\noindent
{\bf 6. The Results.}
We will now apply the above strategy to the computation of $d_\fir$ 
at known critical points $(0,W_c)$ of Anderson models from four 
universality classes shown in Table~\ref{table:1}. 
Empirically chosen overall size cut $L_\text{min}\!=\!20$ will be 
imposed in the analysis since it ensures good scaling properties for 
all models considered.

\begin{figure}
   \vskip -0.03in
   \bc
   \includegraphics[width=0.45\textwidth]{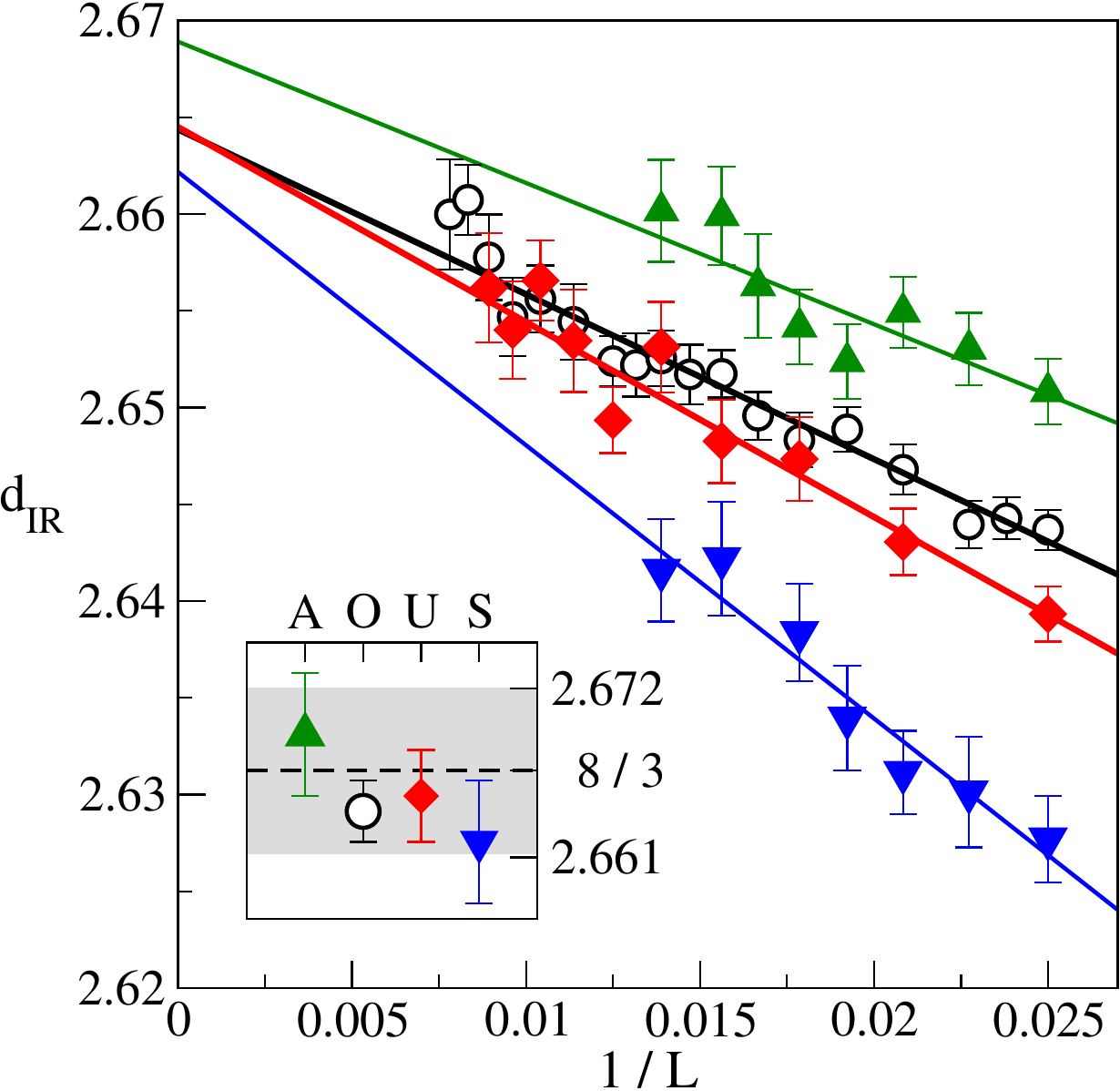}
   \ec
   \vskip -0.15in
   \caption{Dimension $d_\fir$ vs $1/L$ for $L\ge 40$, obtained 
        from Eq.~\eqref{eq:060} with $s\!=\!2$. The inset shows 
        infinite-volume extrapolation for each class. The shaded 
        area marks the band with deviation less than 2 parts per 
        mill from $d_\fir \!=\! 8/3$.}
    \label{fig:2a}
    \vskip -0.17in
\end{figure}

The grand summary of all utilized data is shown in Fig.~\ref{fig:2a}. 
The feature immediately standing out is that dimensions for O and U 
classes become essentially equal in the statistical sense 
for $1/L \!\LessApprox\! 0.02$, approaching together the value 
$d_\fir \!\approx\! 8/3$ in the infinite-volume limit. At the same 
time, S and A dimensions tend to a very similar value in more 
disconnected manner.

The data in Fig.~\ref{fig:2a} leads to $K_\text{O}\!=\!11\,[128]$,
$K_\text{U}\!=\!7\,[8]$, $K_\text{A}\!=\!7\,[1]$ and 
$K_\text{A}\!=\!7\,[1]$,
where the subscript refers to a symmetry class and the bracket specifies
the number of distinct maximal pair combinations. The corresponding
final estimates of $d_\fir$ are given in Table~\ref{table:1}. They
are also shown graphically in the inset of Fig.~\ref{fig:2a}.
In order to represent these final answers faithfully, the straight 
lines for O and U in Fig.~\ref{fig:2a} are the averages from 
fits over all maximal combinations. The standard deviation in 
the associated population of (correlated) estimates is smaller than 
the statistical error by about a factor of two in both cases. 
This confirms the robustness of the method used to obtain 
the extrapolated~$d_\fir$.

\smallskip
\noindent
{\bf 7. The Discussion.}
The properties of Anderson critical points, as expressed by the index 
$\nu$, are known to vary by as much as tens of percents
(Table~\ref{table:1}). This is to be contrasted with our results for 
the effective spatial dimensions $d_\fir$ which differ at the level 
of couple parts per mill at most. In fact, the remaining statistical 
and mild systematic (entering via $L_\text{min}$) uncertainties open 
the possibility that  $d_\fir$ may be strictly {\em super-universal}, 
taking the value of approximately $8/3$ at Anderson transitions.

The relevance of critical $d_\fir$ is that it describes the geometry
of space associated with Anderson transition in the same way as 
Minkowski or Hausdorff dimensions describe the geometry of subsets
in Euclidean space: it is a dimension based on measurable physical
volumes. Its meaning can be illustrated by a fictional inquiry 
about the properties of space addressed to Anderson $E\!=\!0$ 
electrons. The response of O-electrons may read: 
``Our probing means are limited but this is what we can say. 
If space is sprinkled with disorder of strength $W \!<\! W_c$, 
then doubling the lengths in all directions gives us $2^3$ times more 
volume to effectively spread into when these lengths are large. 
Hence, we see space as 3-dimensional. But when disorder of strength 
$W \!>\!W_c$ is used, this factor approaches unity ($2^0$) for large 
lengths, and the space acts 0-dimensional. Most interestingly, 
for $W\!=\!W_c$, our effective volume grows by a factor close 
to $2^{8/3}$ and we have no choice but to tell you that the dimension 
of space we experience in this case is about $8/3$." 

Super-universality of $d_\fir$ conveys that the response from U, S, 
and A-electrons, and possibly others, will be identical to the one 
above. In other words, that the most basic characteristic of spatial 
geometry involved in an Anderson transition, dimension through which 
it proceeds, is insensitive to the symmetries involved. Rather, it is 
entirely determined by the defining attribute of these transitions as 
changes from diffusive to non-diffusive dynamical regimes of quantum 
particles subject to spatial disorder. The proposed super-universal 
status could make $d_\fir$ a generic fingerprint of the Anderson 
phenomenon.   

Our reasoning is made possible by the effective number 
theory~\cite{Horvath:2018aap,Horvath:2018xgx} which gives 
the effective volume based on Eq.~\eqref{eq:020} its measure-like 
character and reveals its absolute meaning. Conversely, the growing 
evidence that the associated $d_\fir$ leads to productive results
(see also~\cite{Alexandru:2021pap}) confirms the usefulness of 
the underlying ideas. In fact, the mathematical basis for $d_\fir$ 
is yet more solid. Indeed, upon formalizing the notion of 
measure-based dimension for probability-induced effective subsets 
it can be shown that $d_\fir$ is the only dimension of this 
type~\cite{Horvath:2022ewv}.


Spatial properties of Anderson transitions have traditionally 
been analyzed using the IPR-based dimensions $D_q$ 
(see e.g.~\cite{Mildenberger:2002a,Vasquez:2008a,Rodriguez:2008a,
Ujfalusi:2015a} or reviews~\cite{Markos:2006A,Evers_2008}). 
Since these dimensions are not measure-based and their rationale is 
different, the information provided by $d_\fir$ is complementary 
to the one accumulated in such studies. Combining inputs from both 
approaches will eventually result in a more complete spatial picture 
of Anderson criticality. 

Important conclusion derived from IPR-based studies is that 
the critical space structure of Anderson transition is much more 
complex than a scale invariant fractal. The question then arises
what this means for the measure-based approach. The answer is 
built into standard notions of dimension such as Hausdorff, 
Minkowski or topological, as well as into $d_\fir$. In particular, 
for dimensional composites, all these concepts select out the maximal 
dimension present in the structure.

The relevance of $\efNm$-defined effective volumes and $d_\fir$ 
is generic: they characterize the geometry of space in which 
a physical process occurs. In the context of Anderson localization, 
we expect a direct connection to the effect of anomalous critical 
diffusion~\cite{Sheng:2006a}. Indeed, this is believed to arise due 
to the process being restricted to a subvolume of the sample effectively 
occupied by the critical electron~\cite{Alexander:1982a}. Hence, 
the measure-based approach is appropriate for its physics. Elementary 
scaling arguments suggest that the diffusion exponent doesn't depend 
on symmetries~\cite{Sheng:2006a}, which would conform to the suggestion 
that the effect descends from super-universal geometry of the subvolume.

Among motivations leading to $d_\fir$ was a need for such characteristic 
in studies of Dirac modes in Quantum Chromodynamics (QCD). One recent 
novelty in that area is a power singularity of mode density at eigenvalue 
$\lambda_\fir \!=\! 0$, appearing in thermal QCD at certain temperature. 
Its existence sparked the proposal for a new scale-invariant phase 
of strongly interacting matter~\cite{Alexandru:2019gdm}. The vicinity of 
$\lambda_\fir$ was shown to have certain properties normally associated 
with localization~\cite{Alexandru:2021pap,Alexandru:2021xoi}, suggesting 
that it can be viewed as a critical point of Anderson type. In addition, 
there is a known Anderson-like point $\lambda_\text{A}$ in the bulk 
of the spectrum~\cite{Kovacs:2010wx, Giordano:2013taa,Ujfalusi:2015nha}.
These developments raise interesting questions about the degree of 
similarity between such QCD features, arising from complicated gauge 
dynamics, and pure Anderson transitions. The results presented here and 
further studies of $d_\fir$ in both contexts will likely help to resolve 
such questions. 

\medskip

\begin{acknowledgments}
P.M. was supported by Slovak Grant Agency VEGA, Project n. 1/0101/20.
I.H. acknowledges the discussions with Andrei Alexandru and Robert Mendris. 
\end{acknowledgments}


\bibliography{my-references}

\end{document}